\documentclass[10pt,letterpaper]{article}

\usepackage{opex3}
\usepackage{graphicx}
\usepackage{color}
\usepackage{amsmath}
\usepackage{here}

\newcommand{\be}{\begin{equation}}
\newcommand{\ee}{\end{equation}}
\newcommand{\bea}{\begin{eqnarray}}
\newcommand{\eea}{\end{eqnarray}}

\newcounter{Fig}

\begin{document}

\begin{sloppy}

\title{Nonlinear plasmonic slot waveguides}

\author{Arthur R. Davoyan, Ilya V. Shadrivov, and Yuri S. Kivshar}

\address{Nonlinear Physics Center, Research School of Physical
Sciences and Engineering, \\Australian National University, Canberra
ACT 0200, Australia }

\email{ard124@rsphysse.anu.edu.au}

\begin{abstract}
We study nonlinear modes in subwavelength slot waveguides created by a nonlinear dielectric slab sandwiched between two metals. We present the dispersion diagrams of the families of nonlinear plasmonic modes and reveal that the symmetric mode undergoes the symmetry-breaking bifurcation with the energy primarily localized near one of the interfaces. We also find that the antisymmetric mode may split into two brunches giving birth to two families of nonlinear antisymmetric modes.
\end{abstract}



\section{Introduction}

Recent progress in nanofabrication opens novel opportunities for engineering smaller optical devices. One of the rapidly emerging fields is the study and design of optical integrated circuits, which would allow increasing functionalities of basic operating elements in information processing. However there exist some fundamental limits for scaling of optical elements and their overcoming is a challenging physical and engineering problem. It is believed that incorporating metals, compatible with nowadays electronics, into optical elements would allow overcoming fundamental diffraction limits by surface plasmon excitation, squeezing light at subwavelength scale, and increasing nonlinear response for switching and signal processing applications. Thus plasmonic waveguides and elements are of great interest these days.

Some basic linear plasmonic elements and devices have been proposed recently~\cite{Bozh}. One of the simplest plasmonic waveguides is an interface between metal and insulator supporting plasmon polaritons; however, due to losses in metal excited plasmon propagates very short distances. Introducing three-layer system helps to increase substantially the propagation distances due to the coupling of plasmons at the neighboring interfaces and the field concentration in dielectric rather than metal~\cite{Maier}. Thus, two possible geometries seem interesting for guiding of plasmons, namely, insulator-metal-insulator and metal-insulator-metal. In past decades rigorous linear analysis of these structures has been presented~\cite{Economou,Burke,Prade}. However, to the best of our knowledge, nonlinear waveguides of this type have not been analyzed yet. We can mention only a few papers concerning the study of nonlinear plasmons at the metal-dielectric interface where the dielectric possesses the Kerr nonlinear response~\cite{Agranovich, Stegeman, Mihalache, Boardman}.

Since only TM electromagnetic waves can be supported by a metal-dielectric interface, the corresponding nonlinear Maxwell equations involve two components of the electric field, and they can't be solvable analytically in a general case and several approximations and simplifications have been employed. The first, uniaxial approximation, proposed by Agranovich {\em et al.}~\cite{Agranovich}, allows to solve Maxwell's equations when nonlinearity depends on the longitudinal component of the electric field. However, in real physical systems the longitudinal component is weaker than the transverse component and, therefore, this assumption is valid in specific cases only. Stegeman {\em et al.} assumed that nonlinearity is caused only by the transverse field component~\cite{Stegeman}; this approximation has also some limitations. Numerical studies of the nonlinear problem with a single interface has been presented in Refs.~\cite{Mihalache,Boardman} while some other studies~\cite{Leung,Yu} presented the mathematical analysis of the Maxwell equations in quadratures, but the results of this analysis are hard to generalize to multilayered systems.

 Based on the uniaxial approximation~\cite{Stegeman}, the studies of long range plasmon-polaritons in metallic films embedded into nonlinear materials have been  carried out as well~\cite{Ariyasu}. But, to the best of our knowledge, the modes of nonlinear slot waveguides have been not studied yet. In this paper, we provide detailed numerical analysis of the problem of {\em nonlinear plasmonic slot waveguide} calculating all possible modes in the metal-insulator-metal structure including the mode splitting and bifurcations. We demonstrate that the symmetric nonlinear mode undergoes bifurcation via symmetry breaking, and a new asymmetric mode with the energy localized near one of the interfaces emerges. In addition, we demonstrate an interesting effect of the splitting of antisymmetric modes that is associated with the multivalued  solutions for linear guided modes, which were not discussed in earlier literature. This results suggest interesting applications for nonlinear switching of plasmonic guided modes.

The paper is organized as follows. In Sec.~\ref{lin} we formulate our problem and discuss the main equations and dispersion relations for linear systems. In Sec.~\ref{nonlin} we consider the guided wave solutions for the corresponding nonlinear problem. Finally, Sec.~\ref{concl} concludes the paper.

\section{Linear guided modes} \label{lin}

 We analyze the guided modes in a symmetric slot waveguide created by a three-layer structure: metal - insulator - metal. The schematic of the structure is presented in the inset of Fig.~\ref{linear}. For simplicity of the mode analysis, we assume that metal is lossless. This assumption can be justified in the case of low attenuation in the frequency range where losses do not change much the shape of the transverse modes studied below but affect their propagation length.

In such structures only TM modes exist, so that we can present the field components in the form, ${\bf H}={\bf e_y}H_y(x)\exp (-j\beta z)$ and ${\bf E} = [{\bf e_x}E_x(x)+j{\bf e_z}E_z(x)]\exp(-j\beta z)$; also we have considered that $E_z$ is shifted in phase with respect to $E_x$ and $H_y$~\cite{Mihalache}. Maxwell's equations for TM modes can be written in the form,
\bea  \label{Maxwell}
\frac{dH_y}{d x} = -\varepsilon E_z, \;\;\; \beta H_y = \varepsilon E_x, \;\;\;
\frac{d E_z}{d x}+\beta E_x = H_y,
\eea
where $\beta$ is the effective index (propagation constant), and the coordinates $x$ and $z$ are normalized to $2\pi/\lambda$, where $\lambda$ is free space wavelength.

\begin{figure}[t]
\begin{center}
\includegraphics[width=8.5cm]{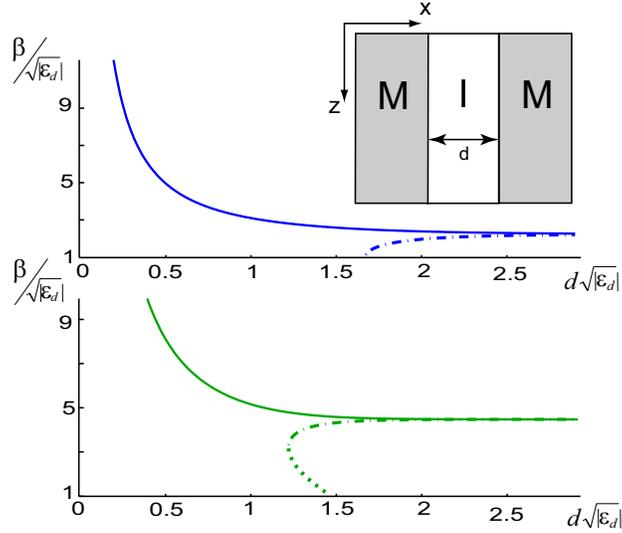}
\end{center}
\caption{ Dispersion of the linear guided modes shown as the dependence of the normalized guided index vs. normalized slot width, for different values of $\rho=|\varepsilon_m|/\varepsilon_d$: $\rho=1.3$ (top) and $\rho=1.05$ (bottom), for symmetric (solid) and two antisymmetric modes (dashed and dotted). Inset shows the schematic of the structure, shaded regions correspond to metal. } \label{linear}
\end{figure}

The variation of the refractive index across the structure is taken in the following form,
\be
   \varepsilon = \left\{ \varepsilon_d, \;\; x \in (0,d) \atop \varepsilon_m, \;\; x \notin (0,d)\right.
\ee
where $d$ is the thickness of the dielectric layer.

Implying the boundary conditions of continuity of the field components $H_y$ and $E_z$ at the interface between metals and dielectric, we derive the well-known dispersion relation~\cite{Burke},
\be \label{dispersion}
\tanh(\lambda_d d)[\varepsilon_m^2\lambda_d^2+\varepsilon_d^2\lambda_m^2]+2\varepsilon_m\varepsilon_d\lambda_m\lambda_d=0
\ee
where $\lambda_m=\sqrt{\beta^2-\varepsilon_m}$ and $\lambda_d=\sqrt{\beta^2-\varepsilon_d}$.

Detailed analysis of this dispersion relation~\cite{Economou,Burke} demonstrate the existence of two types of solutions for guided modes, with respect to the structure of the magnetic field, symmetric and antisymmetric, also refereed in literature as even and odd modes, respectively.

A systematic analysis~\cite{Prade} demonstrated that the mode existence and structure are defined by the values of the ratio $\rho=|\varepsilon_m|/\varepsilon_d$, so that for $\rho \leq 1$ there exists only one (antisymmetric) mode with higher cutoff with respect to the slot width $d$. For $\rho>1$ two modes exist, symmetric, without cutoff, and antisymmetric, with lower cutoff. However, our analysis reveals that at $\rho$ close to 1 and certain slot widths, the lower branch becomes multi-valued so that formally there exist three modes for a fixed value of the slot width, one symmetric and two antisymmetric.

Figure~\ref{linear} presents the modal dispersion as a function of the slot width $d$,  at different values of $\rho$. As was predicted earlier~\cite{Prade}, at $\rho \gg 1$ two modes exist: symmetric (solid curve), and antisymmetric (dashed curve). For $\rho$ approaching 1 we observe three modes, see Fig.~\ref{linear} (bottom), symmetric mode and two antisymmetric modes. One of the antisymmetric modes is more confined to the interfaces and has lower cutoff (dashed curve), the other mode is less confined (dotted curve), and it appears for certain values of the slot width. Since metals are frequency dispersive, we can always find a frequency range for different insulators where all those regimes can be realized.

\begin{figure}[t]
\begin{center}
\includegraphics[width=8cm]{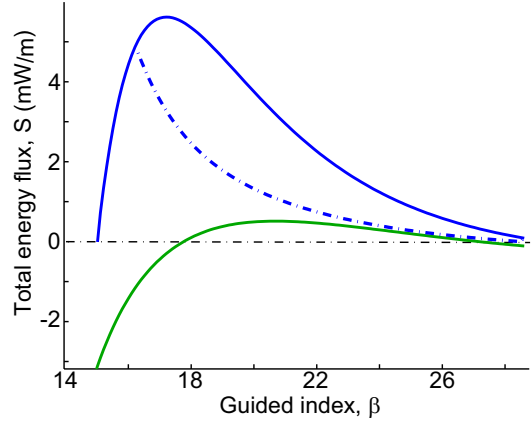}
\end{center}
\caption{ Dispersion of nonlinear guided modes shown as the total power flow vs. guided index for $\lambda=480$nm, $\varepsilon_m=-8.25$, $\varepsilon_d = 7.84$, $\rho=1.05$, $\alpha=1.4\times10^{-18}(m^2/V^2)$, and $d=25$nm. }
\label{case1}
\end{figure}

\section{Nonlinear guided modes}
\label{nonlin}

 We continue our analysis and now assume that the insulator is a nonlinear dielectric with Kerr nonlinear response. To be more specific, we consider that the nonlinear dielectric is chalcogenite glass $As_2Se_3$ with self-focusing nonlinearity sandwiched between two silver slabs. In this case, dielectric permittivity of the nonlinear slab can be presented as follows:
\be \label{nln_perm}
  \varepsilon_{\rm nln}=\varepsilon_d+\alpha (E_x^2+E_z^2)
\ee
where $\varepsilon_d=7.84$, and $\alpha=1.4\times10^{-18}(m^2/V^2)$ is the nonlinear coefficient.

We choose $\rho=1.05$, so that three mode regime is realized, see Fig.~\ref{linear} (bottom), thus $\varepsilon_m=-8.25$. For silver this is realized at $\nu = 0.63 \times 10^{15}Hz$ (free space wavelength $\lambda=480$nm)~\cite{Johnson}.

We solve Eqs. (~\ref{Maxwell}) numerically by applying the numerical shooting method and find the mode profiles for different values of the guided index and nonlinear parameter. To analyze the nonlinear modes we calculate the total energy flux per unit length in the direction of propagation,
\be
  S = \int \left[ \left[{\bf E}\times{\bf H}\right] \times {\bf z} \right] dx.
\ee
and plot this value for three different values of the slot width, corresponding to different types of linear guided waves shown in Fig.~\ref{linear}.

First, we analyze the guided modes when there exists only one symmetric mode in the linear regime.
For this case, we observe that for all values of the guided index $\beta > \sqrt{\varepsilon_d}$ there appears an antisymmetric nonlinear mode, see Fig.~\ref{case1}. Even the total flux for this mode may vanish, this nonlinear mode has no linear analog, and it requires a finite power to be excited in the structure. Our analysis reveals that for smaller slot widths (not shown here) the total energy flux of this antisymmetric mode is negative for all guided indices. With the slot width grow, the total energy flux of this mode becomes positive for some values of $\beta$, see Fig.~\ref{case1}. The typical mode profile is shown in Fig.~\ref{modes}(c).

\begin{figure}[t]
\begin{center}
\includegraphics[width=8.5cm]{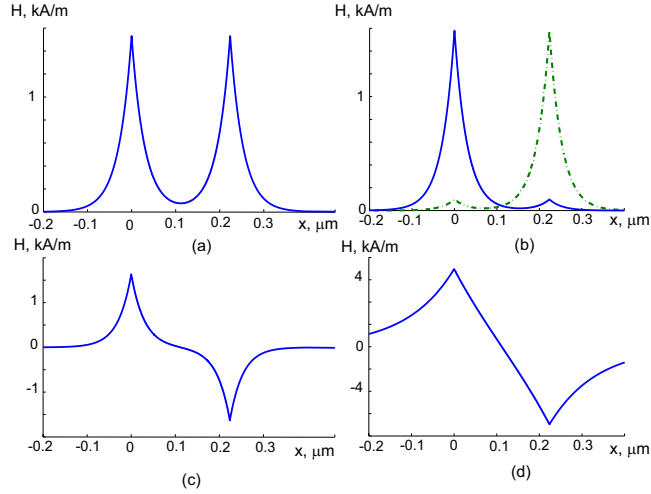}
\end{center}
\caption{ (a-d) Characteristic profiles of nonlinear plasmonic modes shown as the guided magnetic field for different branches of the dispersion curves marked by points in Fig.~\ref{case2}. } \label{modes}
\end{figure}

\begin{figure}[t]
\begin{center}
\includegraphics[width=8cm]{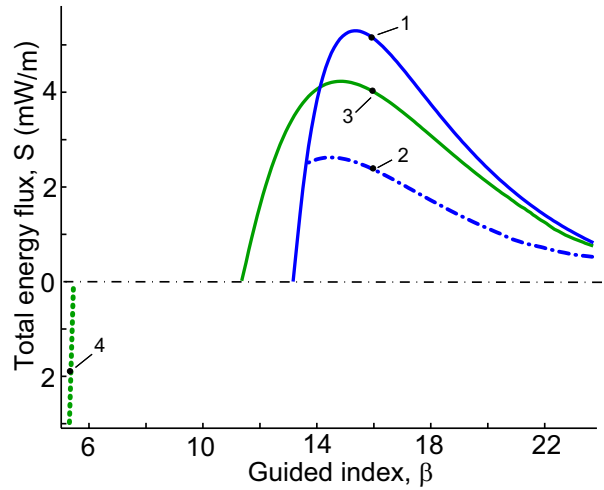}
\end{center}
\caption{Dispersion of nonlinear guided modes shown as the total power flow vs. guided index for $d=35$nm.
In the linear limit of vanishing power, the structure supports one symmetric and two antisymmetric plasmonic modes. } \label{case2}
\end{figure}

For $\beta \simeq 15$, a symmetric mode appears, and for larger $\beta$ the flux increases up to its maximum ($S \simeq 6 mW/m$) at $\beta \simeq 17$ but then decreases again. Also, at $\beta \simeq 16$ we observe a symmetry breaking bifurcation (at $S \simeq 5 mW/m$) which leads to the appearance of an asymmetric mode with power decrease monotonously with $\beta$. For larger $\beta$ the field becomes strongly confined to the interfaces, thus the interaction between plasmons localized at different interfaces becomes weaker, so that the dispersion resembles that of a single interface plasmon~\cite{Boardman}. Typical mode profiles of symmetric and asymmetric modes are presented in Fig.~\ref{modes}(a,b), respectively.

For larger values of the slot width, three modes appear excited and the antisymmetric mode splits into two branches. With larger $d$ (see Fig.~\ref{case2}), the character of the symmetric mode and bifurcated asymmetric branch does not change much. However, the bifurcation point is observed at smaller values of power. Also, the asymmetric mode reaches  its maximum. The mode profiles are shown in Fig.~\ref{modes}(a-d) for the marked points (1-4). Long-wave antisymmetric mode (see Fig.~\ref{case2}, dotted) has a negative flux because the field resides mainly in metal, see Fig.~\ref{modes}(d), and the mode is less confined to the interfaces in comparison to the short-wave antisymmetric mode, [see Fig.~\ref{case2}, dashed; and Fig.~\ref{modes}(c)]. We note that for all  slot widths there is a range of guided indices where several modes at different powers can coexist.

Further increase of the slot width leads to broadening of the gap between the antisymmetric branches bringing to the degeneracy of long-wave antisymmetric branch since it becomes radiative. The bifurcation power decreases, and it is observed at $S \simeq 1 mW/m$, see Fig.~\ref{case3}.

\begin{figure}[t]
\begin{center}
\includegraphics[width=8cm]{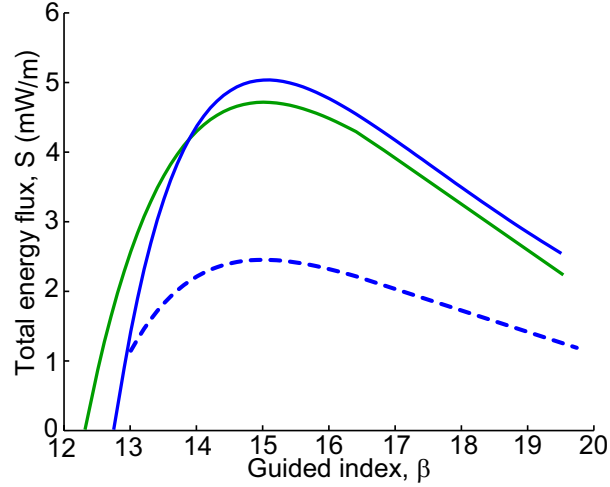}
\end{center}
\caption{Dispersion of nonlinear guided modes shown as the total power flow vs. guided index for $d=50$nm.
In the linear limit of vanishing power, the structure supports one symmetric and one antisymmetric plasmonic modes.} \label{case3}
\end{figure}

\section{Conclusions and acknowledgements}
\label{concl}

We have studied the families of plasmonic modes in nonlinear slot waveguides and predicted the symmetry-breaking bifurcation of the symmetric modes with the critical power depending on the slot width. We have also discussed a complex structure of the asymmetric plasmonic modes that original from the splitting of linear plasmonic modes.

The authors thank  S.I. Bozhevolnyi, D. Gramotnev, R. McPhedran, M. Stockman, and A. Zayats for useful suggestions and acknowledge a support of the Australian Research Council.

\end{sloppy}
\end{document}